\pdfoutput=1
\documentclass[11pt]{article}

\usepackage{graphicx,color}
\usepackage{amsmath}
\usepackage{dcolumn}
\usepackage{bm}
\usepackage{slashed}
\usepackage[utf8]{inputenc}
\usepackage{authblk}

\usepackage{ulem}

\newcommand{\be}{\begin{equation}}
\newcommand{\ee}{\end{equation}}
\newcommand{\ba}{\begin{eqnarray}}
\newcommand{\ea}{\end{eqnarray}}
\newcommand{\no}{\nonumber \\}
\newcommand{\gsim}{\mathrel{\hbox{\rlap{\lower.55ex \hbox {$\sim$}}
                   \kern-.3em \raise.4ex \hbox{$>$}}}}
\newcommand{\lsim}{\mathrel{\hbox{\rlap{\lower.55ex \hbox {$\sim$}}
                   \kern-.3em \raise.4ex \hbox{$<$}}}}

\def\roughly#1{\mathrel{\raise.3ex\hbox{$#1$\kern-.75em%
\lower1ex\hbox{$\sim$}}}}
\def\lsim{\roughly<}
\def\gsim{\roughly>}
\def\fm{{\mbox{fm}}}

\def\({\left(}
\def\){\right)}
\def\[{\left[}
\def\]{\right]}
\def\<{\langle}
\def\>{\rangle}
\def\pd{\partial}

\def\d{{\delta}}
\def\D{{\Delta}}

\def\a{{\alpha}}
\def\b{{\beta}}
\def\c{{\chi}}

\def\g{{\gamma}}
\def\G{{\Gamma}}
\def\p{{\pi}}

\def\m{{\mu}}
\def\n{{\nu}}
\def\r{{\rho}}
\def\s{{\sigma}}

\def\t{{\tau}}

\def\ph{{\phi}}

\def\Ps{{\Psi}}
\def\x{{\xi}}

\def\et{{\eta}}
\def\fm{{\text{fm}}}
\def\cs{{\text{CS}}}
\def\mev{{\text{MeV}}}

\newcommand{\tn}{\tilde{n}_5}
\newcommand{\tx}{\tilde{\xi}}

\newcommand{\lag}{\langle}
\newcommand{\rag}{\rangle}

\title{\bf Axial Charge Fluctuation and Chiral Magnetic Effect from Stochastic Hydrodynamics}

\author[1]{Shu Lin\thanks{linshu8@mail.sysu.edu.cn}}
\author[2]{Li Yan\thanks{li.yan@physics.mcgill.ca}}
\author[1]{Gui-Rong Liang\thanks{bluelgr@sina.com}}
\affil[1]{School of Physics and Astronomy, Sun Yat-Sen University, Zhuhai, 519082, China}
\affil[2]{Department of Physics, McGill University, 3600 rue University, Montr\'eal, QC 
H3A 2T8, Canada }

\date{\today}

\begin{document}

\maketitle

\begin{abstract}
The amount of axial charge 
produced in heavy ion collisions is one of the key quantities in understanding chiral magnetic effect. Current phenomenological studies assume large 
axial charge chemical potential
$\m_5$ produced in Glasma phase and assume the conservation of axial charge throughout the evolution, which is valid in the long relaxation time limit. 
Based on the solution of stochastic hydrodynamics with phenomenological parameters,
our study suggests that the situation of heavy ion collisions may be close to the opposite limit, in which axial charge fluctuation approaches thermodynamic limit. Using $\m_5$ set by the thermodynamic limit for chiral magnetic effect and a background from parity-even $v_1$, we obtain a reasonable description of the centrality dependence of charged particle correlation measured in experiment.
\end{abstract}

\newpage

\section{Introduction}

It is believed that parity (P) and charge-parity (CP) odd domains exist in quark-gluon plasma (QGP) produced by heavy ion collisions (HIC). Such domain may be created by topological fluctuations \cite{McLerran:1990de,Moore:2010jd} (see also \cite{Arnold:1987mh,Arnold:1996dy,Arnold:1998cy} for topological fluctuation in the the context of hot electroweak theory) and quark mass effect \cite{Guo:2016nnq,Hou:2017szz}. In the presence of magnetic field and vorticity, these P odd domains lead to novel phenomena of chiral magnetic effect (CME) \cite{Kharzeev:2007jp} and chiral vortical effect (CVE) \cite{Erdmenger:2008rm,Son:2009tf,Neiman:2010zi,Landsteiner:2011cp}. Both CME and CVE are macroscopic manifestations of chiral anomaly, which are expected 
in off-central heavy ion collisions. There have been active search of CME and CVE in heavy ion collisions 
(cf. \cite{Kharzeev:2015znc,Huang:2015oca,Liao:2014ava}). 
Unfortunately, experimental search for CME and CVE suffers from background such as local charge conservation \cite{Schlichting:2010qia}, transverse momentum conservation \cite{Bzdak:2010fd} and final state interaction \cite{Ma:2011uma}. Different detection techniques \cite{Voloshin:2004vk,Bzdak:2012ia,Wen:2016zic,Xu:2017qfs} have been proposed to reduce these
backgrounds. 

On the phenomenological 
side, one of the largest uncertainties comes from axial charge. Since axial charge is known to be generated from either topological fluctuations or quark mass type fluctuation, it is of stochastic nature. The existing phenomenological studies employ event-by-event anomalous hydrodynamic simulations \cite{Hirono:2014oda,Jiang:2016wve,Shi:2017cpu}. The main ingredients of anomalous hydrodynamics is stochastic initial axial charge generated from flux tube in Glasma phase \cite{Fukushima:2010vw,Mace:2016svc}, and the assumption of axial charge conservation throughout the evolution. However, axial charge is known to dissipate on a time scale given by $\t_\cs=2\c T/\G_{\cs}$, where
$\chi$ is the charge susceptibility and 
$\G_{\cs}$ is the Chern-Simon (CS) diffusion rate\footnote{Here we assume CS diffusion is the main mechanism of axial charge generation}. This is the relaxation time scale. The assumption of axial charge conservation is valid when the time scale of hydrodynamic evolution 
is much less than the relaxation time. It is by no means obvious whether this condition is satisfied 
in the context of heavy ion collisions. 

In fact, the hydrodynamic framework incorporating both axial charge generation and dissipation effect has been written down by one of us \cite{Iatrakis:2015fma}. The hydrodynamic equations for axial charge are given by
\begin{align}\label{stohydro_static}
\left\{\begin{array}{l}
\pd_tn_5+\nabla\cdot {\bf j}_5=-2q, \\
j_{5i}=-D\partial_i n_5+ \xi_i,\qquad i=1,2,3 \\
q=\frac{n_5}{2\t_{\text{CS}}}+\x_q, \\
\end{array}
\right.
\end{align}
The distinct features of Eq \eqref{stohydro_static} is that it contains two types of noises: $\xi_i$ and $\x_q$. The noise $\xi_i$ characterizes thermal fluctuations of conserved charge in a fluid, while
the noise $\x_q$ is unique for the axial charge. 
The noise term $\x_q$ corresponds to the fluctuation of topological charge density $q$, 
which generates total axial charge. The dissipation of the axial charge is controlled
 by the term $\frac{n_5}{2\t_{\text{CS}}}$. The appearance of $q$ (and also
$\x_q$) is a direct consequence of axial charge anomaly, which breaks axial charge conservation.
One thus expects a change of total axial charge from these equations.
The ensemble 
averages of noises are taken to be  
\begin{align}\label{noises}
&\lag\x_q(t,{\bf x})\x_q(t',{\bf x}')\rag=\G_\text{CS}\d(t-t')\d^3({\bf x}-{\bf x}'), \no
&\lag\x_i(t,{\bf x})\x_j(t',{\bf x}')\rag=2\s T\d_{ij}\d(t-t')\d^3({\bf x}-{\bf x}'), \no
&\lag\x_i(t,{\bf x})\x_q(t',{\bf x}')\rag=0,
\end{align}
which relate the strength of flucutations to the Chern-Simons diffusion rate and 
charge conductivity $\sigma$. Correlations of the mixture of these two types of noise vanish.
Eqs. \eqref{stohydro_static} and \eqref{noises} consist stochastic hydrodynamics for axial charge. Note that we assume axial charge is small so that its dynamics decouples from the evolution of energy density and flow. Although Eqs. \eqref{stohydro_static} and \eqref{noises} are written down for a static flow, covariant generalization to arbitrary flows is straightforward.

The aim of this paper is to apply stochastic hydrodynamics to the evolution of axial charge in heavy ion collisions. Accordingly, 
CME signal in experiment is to be quantified. Throughout the paper,  as an approximation and
 simplification to heavy ion collisions, we solve the stochastic hydrodynamics for the axial charge
 evolution on top of Bjorken flow. In Section \ref{sec2}, we first generalize the above stochastic hydrodynamics for arbitrary flows. 
 We then proceed to study axial charge evolution without initial 
 charge density. Our model parameters indicate a relaxation time well within the QGP evolution time. We find the axial charge saturates to thermodynamic limit at late time due to a balance between fluctuation and dissipation effects. In Section \ref{sec3}, we study axial charge evolution with large initial charge density from Glasma phase. In this case, dissipation effect becomes dominant, we find that the axial charge tends to the thermodynamic limit again at late time. In Section \ref{sec4}, we apply the thermodynamic limit of axial charge to the phenomenology of CME. We also introduce a background from fluctuations of parity-even $v_1$. This background has not been discussed in the context of CME before, but it is well constrained by independent 
 flow studies. The CME signal and background combined gives a reasonable description of the
 experimental data, with caveats on the omission of other sources of background and uncertainty. We conclude in Section \ref{sec5}.

\section{Axial Charge Evolution from Vanishing Initial Charge}\label{sec2}

Let us first generalize Eqs. \eqref{stohydro_static} and \eqref{noises} to a covariant form:
\begin{align}\label{stohydro_cov}
\left\{\begin{array}{l}
\nabla_\m J_5^\m=-2q, \\
J_5^\m=n_5u^\m-\s TP^{\m\n}\nabla_\n\(\frac{\m_5}{T}\)+\x^\m,\\
q=\frac{n_5}{2\t_{\text{CS}}}+\x_q, \\
\end{array}
\right.
\end{align}
where $P^{\m\n}=g^{\m\n}+u^\m u^\n$ is the projection operator orthogonal to flow velocity $u^\m$,
and $\nabla_\mu$ stands for covariant derivative.
Throughout this paper, we use most plus signature for $g^{\m\n}$. The noise averages should be generalized accordingly,
\begin{align}\label{noises_cov}
&\lag \x^\m(x) \x^\n(x')\rag=P^{\m\n}2\s T\frac{\d^4(x-x')}{\sqrt{-g}}, \no
&\lag \x_q(x)\x_q(x')\rag=\G_\text{CS}\frac{\d^4(x-x')}{\sqrt{-g}}, \no
&\lag \x^\m(x) \x_q(x')\rag=0.
\end{align}
Note that the thermal noise of conserved charge $\xi^\m$ is generically constructed normal 
to flow four-velocity, which
has only three independent components. 
The parameters appearing in \eqref{noises_cov} are not all independent, but are related by Einstein relations
\begin{align}\label{einstein}
\s=\c D,\quad \t_\text{CS}=\frac{\c T}{2\G_\text{CS}},
\end{align}
with $D$ being the charge diffusion constant. 
Using \eqref{einstein} and also $n_5=\c\m_5$, we can simplify \eqref{stohydro_cov} as
\begin{align}
\nabla_\m\(n_5 u^\m\)-\nabla_\m\(D\c T P^{\m\n}\nabla_\n\(\frac{n_5}{\c T}\)\)+\frac{n_5}{\t_{\cs}}=-\nabla_\m\(P^{\m\n}\x_\n\)-2\x_q.
\end{align}

We use Milne coordinates $x^\m=(\t,\et,x_\perp)$ below. Bjorken flow in Milne coordinate simply gives $u^\m=(1,0,0,0)$, and all the background hydro variables depend only on $\tau$. For instance,
$\chi=\chi(\tau)$.
We assume that the charge densities (including baryon charge, axial charge etc) is small such that their contribution to thermodynamics is negligible. Consequently thermodynamic quantities depend on temperature only, which is a function of $\t$ in Bjorken flow. The only quantity that may depend on other coordinates is $n_5$, because it arises from fluctuations, which can break the Bjorken symmetry.
It follows that
\begin{align}\label{master}
\pd_\t n_5+\frac{n_5}{\t}+\frac{n_5}{\t_\text{CS}}-D\(\pd_\perp^2+\frac{\pd_\et^2}{\t^2}\)n_5=s,
\end{align}
with $s=-\nabla_\m\x^\m-2\x_q$ being source term. Note that $s$ is a linear combination of $\x^\m$ and $\x_q$, thus is of stochastic nature. 
Eq. \eqref{master} is the master equation, solving which gives us the time evolution of axial charge. For pedagogical reason, we solve for the evolution of total axial charge, then we consider rapidity dependence of local axial charge.

\subsection{Total Axial Charge}

The total axial charge $N_5$ is defined as an integration of $n_5$ on hypersurface with constant $\t$.
\begin{align}
N_5=\int d\et d^2x_\perp \t n_5.
\end{align}
In terms of $N_5$, Eq. \eqref{master} takes the following form
\begin{align}\label{N5_eom}
\frac{N_5'}{\t}+\frac{N_5}{\t\t_\text{CS}}=\int d\et d^2x_\perp s.
\end{align}
Here and in what follows, prime indicates derivative with respect to $\tau$.
We claim that the right hand side (RHS) of \eqref{N5_eom} receives contribution from $\x_q$ but not $\x^\m$. The physical reason is $\x^\m$ corresponds to thermal noise, which leads to $n_5$ exchange between fluid cells but leave $N_5$ unchanged. Mathematically, we can show
\begin{align}
\int d\et d^2x_\perp \nabla_\m\x^\m 
=\frac{1}{\t}\int d^3\Sigma \nabla_i\x^i 
=\frac{1}{\t}\int d^2\Sigma_i \x^i=0.
\end{align}
Here $d^3\Sigma=\t d\et d^2x_\perp$ is the differential area of hypersurface $\t=\text{constant}$. $d^2\Sigma_i$ is the differential area of the boundary of the hypersurface. The term $\nabla_\t\x^\t$ vanishes due to the transverseness of $\x^\m$. We used Stokes theorem to express the volume integral as a boundary term, giving a vanishing result at spatial infinity.
We can then solve \eqref{N5_eom} as
\begin{align}\label{N5_s}
N_5(\t_2)h(\t_2)-N_5(\t_1)h(\t_1)=\int_{\t_1}^{\t_2} d\t d\et d^2x_\perp \t h(\t)s(\t,\et,x_\perp),
\end{align}
where $h(\t)$ is solution to the following equation
\begin{align}\label{f_eq}
h(\t)=h'(\t)\t_\text{CS}.
\end{align}
Note that $N_5$ is not conserved due to $\x_q$, which generates net axial charge. 
However, if we turn off topological fluctuation by setting $\G_\cs=0$, and it 
follows from Eq. \eqref{einstein} that $\t_\cs=\infty$, we would have $h(\t)=\text{constant}$. The conservation of $N_5$ is thus obvious from Eq. \eqref{N5_s}.
For a nonvanishing $\G_\cs$, the fluctuation of $N_5$ is given by
\begin{align}\label{DeltaN_correlator}
&\lag \(N_5(\t_2)h(\t_2)-N_5(\t_1)h(\t_1)\)^2\rag \no
&=\int_{\t_1}^{\t_2} d\t d\t'\int d\et d^2x_\perp d\et' d^2x_\perp' \t\t' h(\t)h(\t')\lag s(\t,\et,x_\perp)s(\t',\et',x_\perp')\rag.
\end{align}
The integrand can be easily obtained from \eqref{noises_cov}:
\begin{align}
\lag s(\t,\et,x_\perp)s(\t',\et',x_\perp')\rag=4\G_\text{CS}\d(\t-\t')\d(\et-\et')\d^2(x_\perp-x_\perp')/\t.
\end{align}
Plugging the above into \eqref{DeltaN_correlator}, we obtain
\begin{align}\label{N_correlator}
\lag \(N_5(\t_2)h(\t_2)-N_5(\t_1)h(\t_1)\)^2\rag=\int_{\t_1}^{\t_2}d\t d\et d^2x_\perp \t h(\t)^2 4\G_\text{CS}.
\end{align}

To proceed, we need to solve \eqref{f_eq} for $h(\t)$. We assume the transport coefficients scale with temperature as in conformal theory. Dimensional arguments lead to 
$\t_\text{CS}\sim\frac{1}{T}$ and $\G_\text{CS}\sim T^4$. In Bjorken flow, we have $T\sim \t^{-1/3}$. We adopt the following parametrization:
\begin{align}\label{para1}
T=T_0\(\frac{\t}{\t_0}\)^{-1/3},\quad \t_\text{CS}=\(\frac{\t}{\t_0}\)^{1/3}\t_{\text{CS}0},\quad \G_\text{CS}=\G_0\(\frac{\t}{\t_0}\)^{-4/3}.
\end{align}
The parameter 
$\t_0$ is an arbitrary time scale, which we take to be the initial time of hydrodynamics $\t_0=0.6$ fm. Parameters with index $0$ correspond to their values at $\t=\t_0$, or equivalently $T=T_0$. Using Eq. \eqref{para1}, we can solve Eq. \eqref{f_eq} as
\begin{align}\label{f_sol}
h(\t)=e^{\frac{3}{2}\(\frac{\t}{\t_0}\)^{2/3}}\(\frac{\t_0}{\t_{\cs0}}\).
\end{align}
We have fixed normalization of $h(\t)$, which does not affect the results. 

Let us further assume vanishing initial axial charge density at the beginning of hydrodynamic evolution: $N_5(\t_1=\t_0)=0$. We obtain upon plugging \eqref{f_sol} into \eqref{N_correlator}
\begin{align}\label{N_final}
\lag\(N_5(\t_2)^2\)\rag=\int d\et d^2x_\perp 2\G_0\t_0\t_{\cs0}\(1-e^{3\(1-\(\frac{\t_2}{\t_0}\)^{2/3}\)\(\frac{\t_0}{\t_{\cs0}}\)}\).
\end{align}
At early time $\t_2,\,\t_0\ll\t_{\cs0}$, \eqref{N_final} simplifies to
\begin{align}\label{N_final_early}
\lag\(N_5(\t_2)^2\)\rag=\int d\et d^2x_\perp 6\G_0\t_0^2\(\(\frac{\t_2}{\t_0}\)^{2/3}-1\).
\end{align}
Rather than the random walk (linear) growth, the growth is $\t^{2/3}$ in Bjorken flow, indicating a slower growth due to longitudinal expansion. At late time $\t_2\gg \t_0,\,\t_{\cs0}$, \eqref{N_final} reduces to
\begin{align}\label{N_final_late}
\lag\(N_5(\t_2)^2\)\rag=\int d\et d^2x_\perp 2\G_0\t_0\t_{\cs0}=\int \t_0d\et d^2x_\perp \c_0T_0.
\end{align}
In the last step we have used Einstein relation at $\t=\t_0$, with $\c_0$ being the susceptibility at $\t=\t_0$.
This is $\t_2$ independent, to be compared with $\lag N_5^2\rag=\c TV$ in static case. 
It might be surprising that the late time result \eqref{N_final_late} has the same form of $\c  TV$ at $\t=\t_0$. In fact, we can show that it is also independent of choice of initial time $\t_0$. To verify that, we need to use the scalings: $\G_0\sim \t_0^{-4/3}$ and $\t_{\cs0}\sim\t_0^{1/3}$. Similarly, we can also verify the $\t_0$ independence of \eqref{N_final_early}.

\subsection{Rapidity Dependence}\label{sec_rapidity}

Now we proceed to local fluctuations of axial charge. 
We have two additional transports: conductivity $\s$ and diffusion constant $D$.
Using similar dimensional analysis for the scaling of diffusion constant $\s\sim T\sim \t^{-1/3}$ and $D\sim 1/T\sim \t^{1/3}$, we adopt 
\begin{align}
\s=\s_0\(\frac{\t}{\t_0}\)^{-1/3},\quad D=D_0\(\frac{\t}{\t_0}\)^{1/3}.
\end{align}
To solve for local fluctuation, we start with Fourier transform $\int d\et d^2x_\perp e^{ik_\et\et+ik_\perp x_\perp}$ of \eqref{master}:
\begin{align}\label{master_k}
\frac{\tn}{\t}+\frac{\tn}{\t_\cs}+\tn'+D\(k_\perp^2+\frac{k_\et^2}{\t^2}\)\tn=ik_\et\tx^\et+ik_\perp\tx^\perp-2\tx_q.
\end{align}
We use tilde to indicate Fourier transformed quantities such as $\tn=\int d\et d^2x_\perp e^{ik_\et\et+ik_\perp x_\perp}n_5$. We still denote the RHS of \eqref{master_k} by $\tilde{s}$ and define
\begin{align}\label{f_eq_k}
\frac{h'}{h}=D\(k_\perp^2+\frac{k_\et^2}{\t^2}\)+\frac{1}{\t_\cs}.
\end{align}
With $h$, $\tn$ can be solved similarly as
\begin{align}
\t_2 \tilde n_5(\t_2)h(\t_2)-\t_1\tilde n_5(\t_1)h(\t_1)=\int_{\t_1}^{\t_2} d\t\t h(\t)\tilde{s}(\t).
\end{align}
We need the Fourier transform of noise averages \eqref{noises_cov}
\begin{align}
&\lag \x^\et(\t,k_\et,k_\perp)\x^\et(\t',k_\et',k_\perp')\rag=(2\pi)^3\frac{2\s T}{\t^3}\d(\t-\t')\d(k_\et+k_\et')\d^2(k_\perp+k_\perp'), \no
&\lag \x^{i}(\t,k_\et,k_\perp)\x^{j}(\t',k_\et',k_\perp')\rag=(2\pi)^3\d_{ij}\frac{2\s T}{\t}\d(\t-\t')\d(k_\et+k_\et')\d^2(k_\perp+k_\perp'), \no
&\lag\x_q(\t,k_\et,k_\perp)\x_q(\t,k_\et,k_\perp)\rag=(2\pi)^3\frac{4\G_\cs}{\t}\d(\t-\t')\d(k_\et+k_\et')\d^2(k_\perp+k_\perp'),
\end{align}
which combine to give
\begin{align}\label{s_avg}
\frac{\lag s(\t,k_\et,k_\perp)s(\t',k_\et',k_\perp')\rag}{(2\pi)^3}=\(\frac{2\s T}{\t}\(\frac{k_\et^2}{\t^2}+k_\perp^2\)+\frac{4\G_\cs}{\t}\)\d(\t-\t')\d(k_\et+k_\et')\d^2(k_\perp+k_\perp').
\end{align}
On the other hand, \eqref{f_eq_k} can be solved as
\begin{align}
\ln h(\t,k)=\int_{\t_1}^{\t} d\t'\(D(k_\perp^2+\frac{k_\et^2}{\t'{}^2})+\frac{1}{\t_\cs}\).
\end{align}
Note that we fix the normalization of $h$ by setting the lower bound of the integral to $\t_1$.

Plugging the parametrization of transports, we obtain an explicit factor to be used below
\begin{align}\label{f_sol_k}
\frac{h(\t,k)}{h(\t_2,k)}=e^{-c-a k_\perp^2-b k_\et^2},
\end{align}
with
\begin{align}\label{abc}
&a=\frac{3\(-\t^{4/3}+\t_2^{4/3}\)D_0}{4\t_0^{1/3}}, \no
&b=\frac{3D_0}{2\t_0^{1/3}}\(\frac{1}{\t^{2/3}}-\frac{1}{\t_2^{2/3}}\), \no
&c=\frac{3}{2\t_0^{2/3}}\(-\t^{2/3}+\t_2^{2/3}\)\(\frac{\t_0}{\t_{\cs0}}\).
\end{align}
Plugging \eqref{s_avg} and \eqref{f_sol_k} and assuming vanishing initial axial charge $\tn(\t_1)=0$, we obtain local axial charge fluctuation as
\begin{align}
\lag \t_2\tn(\t_2,k)\t_2\tn(\t_2,k')\rag=\int_{\t_1}^{\t_2}d\t e^{-2(c+a k_\perp^2+b k_\et^2)}\(A k_\perp^2+B k_\et^2+C\)\d(k_\et+k_\et')\d^2(k_\perp+k_\perp'),
\end{align}
with
\begin{align}\label{abc2}
&A=2\s_0T_0\t_0\(\frac{\t}{\t_0}\)^{1/3}, \no
&B=\frac{2\s_0T_0}{\t_0}\(\frac{\t}{\t_0}\)^{-5/3}, \no
&C=4\G_0\t_0\(\frac{\t}{\t_0}\)^{-1/3}.
\end{align}
The parameters \eqref{abc} and \eqref{abc2} are in fact related through Einstein relation
\begin{align}
&\frac{\pd a}{\pd\(\frac{\t}{\t_0}\)}=-\frac{A}{2\c_0T_0}, \no
&\frac{\pd b}{\pd\(\frac{\t}{\t_0}\)}=-\frac{B}{2\c_0T_0}, \no
&\frac{\pd c}{\pd\(\frac{\t}{\t_0}\)}=-\frac{C}{2\c_0T_0}.
\end{align}
This allows us to simplify the local axial charge fluctuation as
\begin{align}
\lag\t_2\tn(\t_2,k)\t_2\tn(\t_2,k')\rag=\c_0T_0\t_0\int_{\t_1}^{\t_2}d\t \frac{\pd}{\pd \t}e^{-2(c+a k_\perp^2+b k_\et^2)}\d(k_\et+k_\et')\d^2(k_\perp+k_\perp').
\end{align}
Fourier transform back to the coordinate space, we obtain
\begin{align}\label{n_correlator}
\lag\t_2n_5(\t_2,\et,x_\perp)\t_2n_5(\t_2,0)\rag=\c_0T_0\t_0\int\frac{dk_\et d^2k_\perp}{(2\pi)^3}e^{-ik_\et\et-ik_\perp x_\perp}\(1-e^{-2(c+a k_\perp^2+b k_\et^2)}\vert_{\t={\t_1}}\).
\end{align}
We are primarily interested in the rapidity dependence of axial charge fluctuations. We simply integrate over transverse plane and complete the Gaussian integral of $k_\et$ to obtain
\begin{align}\label{n_eta}
\int d^2x_\perp\lag\t_2n_5(\t_2,\et,x_\perp)\t_2n_5(\t_2,0)\rag&=\c_0T_0\t_0\int\frac{dk_\et}{2\pi}e^{-ik_\et\et}\(1-e^{-2(c+b k_\et^2)}\vert_{\t={\t_1}}\) \no
&=\c_0T_0\t_0\(\d(\et)-e^{-2c}\frac{e^{-\et^2/(8b)}}{2\sqrt{2\pi b}}\vert_{\t=\t_1}\).
\end{align}
Note that the first term in \eqref{n_eta} is a positive delta function, meaning that it is localized in one
fluid cell. The second term is negative, corresponding to anti-correlation over rapidity. As $\t_2\to\infty$ for fixed $\t_1$, $c\to\infty$ but $b$ remains finite. It implies that axial charge remains correlated over rapidity but the magnitude tends to zero.
Integrating over rapidity, we recover the result of total axial charge fluctuation \eqref{N_final}.

Before moving on, we notice that the similar structure is obtained in baryon correlation \cite{Ling:2013ksb,Kapusta:2017hfi}. It was argued that the delta function term corresponds to self-correlation of particles, thus should be excluded in the calculation of balance function. In case of CME, we think it should be kept. The reasoning is clearest in the late time limit when the second term can be ignored. The first term should be interpreted as a measure of amount of charge in fluid cells. Note that we obtained before the fluctuation of total charge $\lag N_5^2\rag=\c TV$, with $V$ the volume containing $N_5$. The first term is just analog of it within a given cell. To see that, we reorganize \eqref{n_eta} in the late time limit
\begin{align}
\lag n_5(\t_2,\et,x_\perp)n_5(\t_2,0)\rag\simeq\frac{\c_0T_0\t_0}{\t_2}\frac{\d(\et)}{\t_2\int d^2x_\perp}.
\end{align}
It is easy to show the first factor is nothing but $\c(\t_2)T(\t_2)$, while the second factor is volume of fluid cell at $\t_2$, with $\d(\et)$ setting size of the cell in rapidity direction. Therefore it contains more than just self-correlation.

Now let us estimate the effective axial chemical potential $\m_5$ relevant for heavy ion collisions. The following parameters will be used: $\t_0=0.6$ fm, $T_0=350$ MeV, corresponding to the Au+Au collisions
at the highest RHIC energy. We also take $\G_0=30\a_s^4T_0^4$ from extrapolation of weak coupling calculation \cite{Moore:2010jd} and $\c_0=3T_0^2$ from free theory result for three flavors. For the coupling constant, we use $\a_s=0.3$. These combine to give $\t_{\cs0}=\frac{\c_0T_0}{2\G_0}\simeq 2.3$ fm. Clearly the condition that relaxation time scale is much larger than hydrodynamic evolution time is violated, given these parameters. 
The late time limit is more appropriate. To fix the volume factor, we assume the axial charge is generated in a ``cylinder'' with transverse size $A_\text{overlap}$ and rapidity span $\D\et$. The transverse profile is centrality dependent. We approximate it by a disk with effective radius $R$ determined by the overlap area $\p R^2=A_\text{overlap}$. The centrality dependence is taken from Table II of \cite{Abelev:2008ab}. We list centrality dependence of $R$ and also number of collisions $N_\text{coll}$ in Table \ref{tab:cen_R} for later reference. 
\begin{table}
\caption{\label{tab:cen_R}The centrality dependence of effective radius $R$.}
\begin{tabular}{ccccccccc}
\text{Centrality}& 60-70\%& 50-60\%& 40-50\%& 30-40\%& 20-30\%& 10-20\%& 5-10\%& 0-5\%\\
\hline
$R(\fm)$& 2.94& 3.51& 4.07& 4.64& 5.24& 5.91& 6.51& 7.00\\
\hline
$N_\text{coll}$& 32.4& 66.8& 127& 221& 365& 577& 805& 1012\\
\end{tabular}
\end{table}
$\D\et$ is determined by the correlation length in rapidity. We set the correlation length by the standard deviation of the Gaussian term in \eqref{n_eta}: $2\D\et^2=8b$. In the limit $\t_2\to\infty$, $\t\to\t_0$, we obtain from \eqref{abc}
$\D\et=\sqrt{\frac{6D_0}{\t_0}}$, with $D_0$ being the diffusion constant at temperature $T_0$. We fix it by the Einstein relation $D_0=\frac{\s_0}{\c_0}$. Here $\s_0$ is the axial charge conductivity, which can be related to electric conductivity $\s_\text{e}$ at the same temperature by
\begin{align}
\frac{\s_0}{\s_\text{e}}=\frac{N_f}{\sum_fe^2q_f^2}
\end{align}
Taking $\s_e\simeq 0.5\sum_fe^2q_f^2T_0$ from lattice measurement \cite{Ding:2010ga}, we obtain $\s_0\simeq 1.5T_0$.
The corresponding total axial charge $N_5$ and $\m_5$ is estimated using \eqref{N_final} as
\begin{align}\label{N_pheno}
&N_5\sim\(\int d\et d^2x_\perp 2\G_0\t_0\t_{\cs0}\)^{1/2}=\(\p R^2\D\et 2\G_0\t_0\t_{\cs0}\)^{1/2}, \no
&\m_5=\frac{N_5}{\p R^2\t\D\et\c}.
\end{align}
The effective $\m_5$ changes with $\t$ as the denominator indicates: the dependence is $\m_5(\t)=\m_5(\t_0) \(\t/\t_0\)^{-1/3}$. We list the centrality dependence of $\m_5(\t_0)$ in Table \ref{tab:cen_m5}.
\begin{table}
\caption{\label{tab:cen_m5}The centrality dependence of $\m_5(\t_0)$.}
\begin{tabular}{ccccccccc}
\text{Centrality}& 60-70\%& 50-60\%& 40-50\%& 30-40\%& 20-30\%& 10-20\%& 5-10\%& 0-5\%\\
\hline
$\m_5(\mev)$& 16.3& 13.7& 11.8& 10.4& 9.17& 8.12& 7.37& 6.86\\
\end{tabular}
\end{table}
Clearly axial charge fluctuation is more significant in peripheral collisions than in central collisions. This is a reflection of the simple fact that fluctuation is suppressed by volume factor.

\section{Axial Charge Evolution from Nonvanishing Initial Charge}\label{sec3}

The calculation in the previous section assumes vanishing initial axial charge density. However, it is known that large initial axial charge density is generated by chromo flux tube in the Glasma phase \cite{Fukushima:2010vw}. In fact, previous phenomenological studies rely on this initial charge as a main source. In this section, we will assess the role of initial charge over the evolution. For simplicity, we only
discuss 
total axial charge. We start with the following equation
\begin{align}
N_5(\t_2)h(\t_2)=N_5(\t_1)h(\t_1)+\int_{\t_1}^{\t_2} d\t d\et d^2x_\perp \t h(\t)s(\t,\et,x_\perp).
\end{align}
Here $N_5(\t_1)$ is the total initial charge generated from the flux tube and last term comes from topological fluctuations. We assume correlation of $N_5(\t_1)$ and $s$ vanishes due to their independent origins. It follows that
\begin{align}
\lag\(N_5(\t_2)h(\t_2)\)^2\rag=\lag\(N_5(\t_1)h(\t_1)\)^2\rag+\int_{\t_1}^{\t_2}d\t d\et d^2\x_\perp\t h(\t)^2 4\G_\cs.
\end{align}
The only modification to \eqref{N_correlator} is the appearance of $N_5(\t_1)$ term on the RHS. 
Plugging the explicit expression of $h$, we obtain
\begin{align}\label{N_glasma}
\lag N_5(\t_2)^2\rag&=\lag N_5(\t_0)^2\rag e^{3\(1-\(\frac{\t_2}{\t_0}\)^{2/3}\)\(\frac{\t_0}{\t_{\cs0}}\)}+\int d\et d^2x_\perp 2\G_0\t_0\t_{\cs0}\(1-e^{3\(1-\(\frac{\t_2}{\t_0}\)^{2/3}\)\(\frac{\t_0}{\t_{\cs0}}\)}\) \no
&=\int d\et d^2x_\perp \c_0T_0\t_0+\(\lag N_5(\t_0)^2\rag-\int d\et d^2x_\perp \c_0T_0\t_0\)e^{3\(1-\(\frac{\t_2}{\t_0}\)^{2/3}\)\(\frac{\t_0}{\t_{\cs0}}\)},
\end{align}
where we have identified $\t_1=\t_0$.
We also split the result into a thermodynamic limit and an exponentially suppressed term. Clearly the initial charge contribution is suppressed at late time.
At early time, we have
\begin{align}\label{N_glasma_early}
\lag N_5(\t_2)^2\rag=\lag N_5(\t_0)^2\rag+\int d\et d^2x_\perp 6\G_0\t_0\t_{\cs0}\(\(\frac{\t_2}{\t_0}\)^{2/3}-1\)\(\frac{\t_0}{\t_{\cs0}}\).
\end{align}
It is simply \eqref{N_final_early} plus initial charge fluctuation.
As an application, we estimate the time when axial charge from topological fluctuation becomes comparable with initial contribution. Following \cite{Jiang:2016wve,Fukushima:2010vw}, we take the parametrization of initial charge 
\begin{align}
\sqrt{\lag n_5(\t_0)^2\rag}\simeq \frac{Q_s^4(\pi \r_{\text{tube}}^2\t_0)\sqrt{N_{\text{coll}}}}{16\pi^2A_\text{overlap}},
\end{align}
where $\r_\text{tube}\simeq 1\fm$ is the transverse size of glasma flux tube, $Q_s$ is the saturation scale taken to be $Q_s\simeq 1\text{GeV}$. $N_{\text{coll}}$ and $A_\text{overlap}$ are taken from Table \ref{tab:cen_R}. With all these, we obtain $\m_5(\t_0)\simeq 35\mev$, which is rather insensitive to centrality as opposed to the equilibrium scenario. 
We can determine the proper time when axial charge from topological fluctuation becomes comparable with initial contribution. It occurs at $\t\simeq 5.0\fm$ for $60-70\%$ centrality and $\t\simeq 9.8\fm$ for $0-5\%$ centrality. Of course, this estimate is based on a discontinuous gluing of Glasma phase and QGP phase. It is known from real time simulations that the Glasma phase has larger $\G_\cs$ \cite{Mace:2016svc}, thus we might expect shorter relaxation time. In any case, the relaxation time scale is quite comparable to the QGP evolution time.

\section{Chiral Magnetic Effect from Equilibrated Axial Charge Fluctuations}\label{sec4}

We have seen in the previous section that axial charge generated during QGP evolution can be as important as initial charge contribution. This contribution has been ignored so far in phenomenological studies of CME. We wish to quantify it in this section. Conventional hydrodynamic models set in when QGP thermalizes, meaning that energy density and charge densities reach their equilibrium values at $\t=\t_0$. We assume axial charge fluctuation also reaches thermodynamic limit at the same time. This could be an over-simplified assumptions. Nevertheless, given the large uncertainty in the axial charge fluctuation and its relaxation time, this assumption allows us to treat dynamics of CME using equilibrium value of the fluctuation, which is under better theoretical control. 
We have estimated the effective chemical potential $\m_5(\t)=\m_5\(\t/\t_0\)^{-1/3}$ with $\m_5(\t_0)$ listed in Table \ref{tab:cen_m5}. This applies to the same volume as before. The CME current in unit of $e$ is given by
\begin{align}
\vec j=C_e\m_5 e \vec B,
\end{align}
where $C_e=\sum_{f}q_f^2\frac{N_c}{2\pi^2}=\frac{1}{\pi^2}$ for three flavors. Considering the fact that magnetic field is orientated out-of- reaction-plane $\Psi_{RP}$ in off-central heavy ion collisions, 
charge separation with respect to 
reaction plane is given by
\begin{align}
Q=C_e\int_{\t_0}^{\t_f} d\t \t d\et 2R \m_5B,
\end{align}
where $\int \t d\et 2R$ denotes the cross section of the CME current with the reaction plane. The integration is from initial time $\t_0$ to freezeout time $\t_f$. The amount of electric charge asymmetry created gives rise to the electric chemical potential
\begin{align}\label{mu_e}
\m_e(\t_f)=\frac{Q}{V\c_Q}.
\end{align}
where $V=\pi R^2\t_f/2\int d\et$ is the volume of QGP above or below the reaction plane at freezeout time. Note that $\int d\et$ cancels out in the numerator and denominator. Note also that the induced electric chemical potential has an asymmetric distribution
in space
following the direction of magnetic field, which we assume as $\sin(\varphi-\Psi_{RP})$
with respect to the reaction plane.
The parameter $\c_Q$ is the electric charge susceptibility, not to be confused with axial charge susceptibility $\c$. We use $\t_f=7\fm$ and $\c_Q=\sum_{f}q_f^2T^2=\frac{2}{3}T^2$ as in the free theory\footnote{Note that the constant $e^2$ is absent. This is because we choose to work with $Q$ in unit of $e$ and $\m_e$ in unit of $1/e$.}.
To simplify the calculation, we use the following form for the magnitude of the 
magnetic field: $B=B_0 e^{-\t/\t_B}$, with $eB_0=10m_{\pi}^2$ and $\t_B=3\fm$. It is homogeneous in transverse plane and rapidity span. 
Putting things together, we obtain a $e\m_e$ for different centralities in Table~\ref{tab:cen_m}.
\begin{table}
\caption{\label{tab:cen_m}The centrality dependence of $e\m_e(\t_f)$.}
\begin{tabular}{ccccccccc}
\text{Centrality}& 60-70\%& 50-60\%& 40-50\%& 30-40\%& 20-30\%& 10-20\%& 5-10\%& 0-5\%\\
\hline
$e\m_e(\t_f)(\mev)$& 4.33& 3.03& 2.26& 1.74& 1.37& 1.07& 0.88& 0.76\\
\end{tabular}
\end{table}
Since we assume an overall neutral QGP, this gives the following electric charge fluctuation
\begin{align}
\lag \m_e(\t_f)\rag =0,\qquad
\lag \m_e(\t_f) \m_e(\t_f)\rag\simeq \m_e(\t_f)^2.
\end{align}

To convert to two-particle correlation, we need to do Cooper-Frye freezeout procedure~\cite{Cooper:1974mv}. 
This gives rise to the spectrum of the generated charged particles,
\begin{align}\label{CF}
\frac{d N_Q^i}{d \phi}=\frac{g_i}{(2\pi)^3}\int\! dy p_\perp dp_\perp\int\! d\s_\m p^\m f_i(x,p),
\end{align}
where $f_i=e^{p_\m u^\m/T_f+Q \m_e/T_f+\m_i/T_f}$ is the phase-space distribution
of the i-th particle species in Boltzmann approximation. 
As a good approximation, we only consider pions and kaons in our
calculations with respect to heavy-ion collisions. Therefore, apart from $\m_e$  we introduce also the 
chemical potential $\m_\p\simeq 80\mev$ 
for pions and $\m_K\simeq 180\mev$ for kaons, regarding a freeze-out temperature
$T_f=T_0\(\frac{\t_f}{\t_0}\)^{-1/3}\simeq 154\mev$~\cite{Teaney:2002aj}. The parameter 
$Q$ is the charge of particles in unit of $e$. The degeneracy factor $g$ is taken to be
$g=1$ for $\p_\pm$ and $K_\pm$ respectively. The integration domain for pseudo-rapidity 
should be taken according to experiments. 

In the case of Bjorken flow, we can simplify the freeze-out integral in 
\eqref{CF} so that the charged particle azimuthal
distribution becomes
\begin{align}
\label{eq:dndp0}
\frac{dN_Q^i}{d\phi}=\frac{g_i}{(2\pi)^3}\int\! dy dm_\perp m_\perp^2 
\int\! \t_f d\et d^2x_\perp \cosh(\et-y) f_i(x,p),
\end{align}
with $m_\perp=\sqrt{p_\perp^2+m^2}$ and the lower bound of $m_\perp$ integration being the rest mass of corresponding meson.  
An integration over the azimuthal angle in Eq~\eqref{eq:dndp0} gives rise to the total yields of the
charged particles, $\langle N_Q^i\rangle$. Note that 
$\langle N_Q^i\rangle $ depends only on the background distribution, while
the charge asymmetry is created entirely from CME. Since the effect of charge asymmetry
is much smaller compared to the background (e.g., $\m_e\ll T_f$), 
we have to the lowest order in $\m_e$:
\begin{align}
\label{eq:dndp1}
\d \frac{d N_Q^i}{ d\phi}&=\frac{Qg_i}{(2\pi)^3}\int\! dy m_\perp^2 dm_\perp \int\!
\t_f d\et d^2x_\perp \cosh(\et-y) f_i(\m_e=0) \m_e, 
\end{align}
which characterizes the asymmetric charged particle distribution due to CME.

Up to this point, we have not taken into account the angular dependence of the induced electric
chemical potential, nor the angular dependence of the background charged particles.
In particular, one notices that an expansion in the transverse plane is required in 
Cooper-Frye to 
generate the azimuthal angle dependence of particles, which
however breaks the Bjorken symmetry we have been considering so far. Nonetheless, 
an alternative way, as being inspired from Eqs \eqref{eq:dndp0}
and \eqref{eq:dndp1}, is to assume the follow ansatz for the generated charged single-particle
spectrum, 
\begin{align}
\label{eq:dn_dis1}
\frac{dN_{\pm}}{d\ph}=\frac{d \lag N_{\pm} \rag }{d\phi}+\frac{1}{4}\D_\pm\sin(\ph-\Psi_{RP}),
\end{align}
which generalizes Ref \cite{Kharzeev:2007jp}. Eq \eqref{eq:dn_dis1} 
contains a background contribution to
the angle dependence in $d\lag N_\pm\rag/d\phi$, in addition to the charge-dependent distribution
induced from CME calculation, Eq \eqref{eq:dndp1}. The form of the
charge-dependent distribution can be understood from the asymmetric distribution
of electric chemical potential in space, $\propto\sin(\varphi-\Psi_{RP})$, followed by a 
saddle-point integration in Cooper-Frye to replace $\varphi$ by $\phi$. 
To express $\D_\pm$ in terms of what we have calculated, we further assume that particles detected in the upper half region $0<\ph-\Psi_{RP}<\pi$ come entirely from QGP above the reaction plane. Similarly particles detected in the lower half region $\pi<\ph-\Psi_{RP}<2\pi$ come entirely from QGP below the reaction plane. It follows that
 \begin{align}\label{dq}
 \D_Q =\sum_{i\in Q}\frac{g_i}{(2\pi)^2}\int\! dy m_\perp^2 dm_\perp \int\!
\t_f d\et d^2x_\perp \cosh(\et-y) f_i(\m_e=0) \m_e(\t_f) ,
 \end{align}
 and also the correlation $\lag \D_\pm^2\rag\sim \lag \m_e(\t_f)^2 \rag$. The multiplicity of charged particles is obtained consistently with Eq~\eqref{eq:dndp0} as
\begin{align}\label{nq}
N_Q=\sum_{i\in Q}\frac{g_i}{(2\pi)^2}\int\! dy dm_\perp m_\perp^2 
\int\! \t_f d\et d^2x_\perp \cosh(\et-y) f_i(x,p,\m_e=0).
\end{align}
In both Eqs~\eqref{dq} and \eqref{nq}, $\int d^2x_\perp=A_\text{overlap}$. It cancels in their ratio. The integration domain of rapidity is taken to be $|\et|<2$. We have also tried different integration domains, finding the ratio almost independent on choice of integration domain.

The background angular distribution $d\lag N_\pm\rag /d\phi$ reflects the charge-independent
evolution of the medium with respect to initial state with event-by-event fluctuations,
whose spectrum can be captured by (normal) viscous hydrodynamics. 
We take a form with Fourier decomposition,
\begin{equation}
\label{eq:dn_dis2}
\frac{d \lag N_\pm\rag }{d\phi} = \frac{\lag N_\pm\rag}{2\pi} 
\left[1+2\sum_{n=1} v_n \cos n(\phi-\Psi_n)\right] ,
\end{equation} 
where the coefficient $v_n$ of the Fourier decomposition defines harmonic flow of order $n$,
while $\Psi_n$ indicates the corresponding participant plane angle.  

To compare with the measured CME signature in experiments, we calculate the following correlator \cite{Voloshin:2004vk} defined according to a correlated two-particle spectrum
\begin{align}
\g_{\a\b}&=\lag\cos(\ph_1^\a+\ph_2^\b-2\Psi_{\text{RP}})\rag ,
\end{align}
with $\a,\,\b=\pm$. 
Given the single-particle 
spectrum in Eq \eqref{eq:dn_dis1}, one is allowed to write the correlated two-particle
spectrum as (for each centrality class),
\begin{align}
\label{eq:tp_dis}
\left\langle\frac{dN^\alpha}{d\phi_1^\alpha}\frac{ dN^\beta}{d\phi_2^\beta}\right\rangle
&=\;
\frac{1}{(2\pi)^2} \lag N^\alpha\rag\lag N^\beta\rag \cr
&+\left[\frac{2 }{(2\pi)^2} \lag N^\alpha\rag\lag N^\beta\rag
  \langle v_1^2\cos2(\Psi_1-\Psi_{RP})\rangle
 -\frac{1}{32}\langle \Delta_\alpha \Delta_\beta\rangle
 \right]\cos(\phi_1^\alpha+\phi_2^\beta-2\Psi_{RP})\cr
 &+\ldots\,,
\end{align}
where terms irrelevant to the $\gamma_{\alpha\beta}$ correlator are suppressed and
contained in ellipsis. Note that there is no cross term stemming from interference between
$d\lag N_\pm\rag /d\phi$ and $\Delta_\pm$, owing to the fact that they are distinct in their origins of 
fluctuations. After integrating over angle with respect to Eq \eqref{eq:tp_dis}, one
obtains the $\gamma_{\alpha\beta}$ correlator,
\begin{align}
\label{eq:gamma1}
\gamma_{\alpha\beta}=
\langle v_1^2\cos2(\Psi_1-\Psi_{RP})\rangle
-\frac{\pi^2}{16}\frac{\langle \Delta_\alpha \Delta_\beta\rangle }{\lag N_\alpha\rag \lag N_\beta\rag} 
= \langle v_1^2\cos2(\Psi_1-\Psi_{RP})\rangle - a_{\alpha\beta} .
\end{align}
In Eq \eqref{eq:gamma1}, the first term on the RHS comes from the background 
flow of $v_1$, which is apparently charge-independent. It characterizes the measured first order
harmonic flow in the reaction-plane.\footnote{ It can be understood as $v_1^2\{\Psi_{RP}\}$.
}
Normally, $v_1$ has a rapidity-even and a rapidity-odd components. However, only the rapidity-even
component contributes to our present calculations in a symmetric rapidity window,
e.g. $|y|<1$ with respect to the measurement carried out by the STAR collaboration \cite{Abelev:2009ad}. 
To
calculate this background contribution, we numerically 
solve viscous hydrodynamics with an input specific
shear viscosity $\eta/s=1/4\pi$, with respect to a lattice equation of state.  For simplicity, we do not
carried out hydrodynamic simulations on an event-by-event basis, but calculate 
the response of $v_1$ to initial dipolar asymmetry for each centrality class at the RHIC energy.
In this way, the event-by-event correlation is determined entirely from initial state fluctuations,
which can be determined from event-by-event simulations of the Monte Carlo Glauber model \cite{Alver:2008aq}.
Note that this correlation is negative \cite{Teaney:2010vd}.
More details of the hydro calculation can be found in Ref \cite{Teaney:2012ke}.

The second term in Eq \eqref{eq:gamma1} is due to the CME and axial charge fluctuations, 
\begin{align}\label{as}
a_{++}=\frac{\pi^2}{16}\frac{\lag\D_{+}^2\rag}{\lag N_+\rag^2},\quad a_{--}=\frac{\pi^2}{16}\frac{\lag\D_{-}^2\rag}{\lag N_-\rag ^2},\quad a_{+-}=\frac{\pi^2}{16}\frac{\lag\D_{+}\D_-\rag}{\lag N_+\rag \lag N_-\rag},
\end{align}
which 
differs for the same and the opposite charged particles by different sign:
$a_{++}=a_{--}=-a_{+-}$.

\begin{figure}
\includegraphics[width=0.85\textwidth]{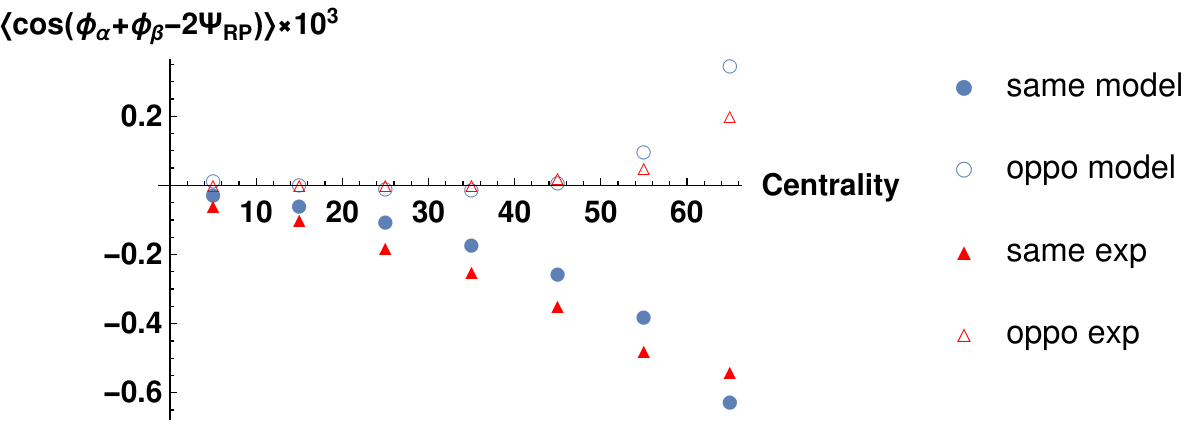}
\caption{\label{cme} Centrality dependence of $\lag\cos(\ph_\a+\ph_\b-2\Ps_\text{RP})\rag\times 10^3$, with circles from model prediction and triangles from experiment \cite{Abelev:2009ad}. Solid symbols are for same sign correlation while empty symbols are for opposite sign correlation.}
\end{figure}

Finally we obtain the centrality dependence of CME signal in Figure \ref{cme}. 
The model calculation yields results as a consequence of 
combined effects from CME and background. Sum of both 
effects gives reasonably well centrality dependence, i.e., the correlations get stronger as in more 
peripheral collisions, in comparison to the measured results. 
In particular, we notice that the background correlation is 
generically negative, which reduces the magnitude of the opposite sign correlation but
increases that of 
the same sign. 
The agreement should be taken with caveat though, as there are many other aspects not included in the model such as background from other sources \cite{Schlichting:2010qia,Bzdak:2010fd} and uncertainty in magnetic field. We leave a more systematic study for future work.
We stress once again that the axial charge used is of different origin from previous studies: it is completely from fluctuations. Since fluctuation is volume suppressed, it gives rise to an effective $\m_5$ with strong centrality dependence, in contrast to weak centrality dependence of $\m_5$ from chromo flux tube contribution.

\section{Conclusion}\label{sec5}

We studied dynamics of axial charge by solving 
stochastic hydrodynamics on top of 
Bjorken flow background. Several interesting results are obtained: in the absence of initial axial charge, we found that expansion slows the growth of $N_5$ at early time, giving $\lag N_5^2\rag\sim\t^{2/3}$ behavior as compared to random walk behavior $\sim \t$ in the non-expanding case. At late time, the total axial charge relaxes to the equilibrium value given by $\lag N_5^2\rag=\c T V$. When large initial axial charge is present, the evolution of axial charge fluctuation consists of exponential relaxation of initial charge and growth of axial charge from topological fluctuation. We found that initial charge relaxes towards 
the thermodynamic limit within the 
time scale of QGP evolution.

We calculated the CME signal assuming that the axial charge fluctuation reaches the thermal equilibrium value at the onset of hydrodynamics. This leads to strong centrality dependence of effective $\m_5$, in contrast to the counterpart from chromo flux tube contribution.
In addition to CME signal, we introduced a background from event-by-event fluctuations of 
parity-even $v_1$.
We found that CME signal from the thermal equilibrium value of axial charge give a reasonable description of experimental data when combined with the rapidity-even $v_1$ background. This indicates that topological fluctuation in QGP can also be a significant source of the axial charge generation. 

Our calculation was done for CME using total axial charge. More differential measurement of CME using particles with fixed rapidity separation is available \cite{Abelev:2009ad}. The calculation can be extended by using local charge fluctuations. We leave refined study for future work.

\section{Acknowledgments}

We are grateful to Jinfeng Liao, Guoliang Ma and Yi Yin for useful discussions. This work is in part supported by One Thousand Talent Program for Young Scholars (S.L.) and NSFC under Grant Nos 11675274 and 11735007 (S.L.), and the Natural Sciences and Engineering Research Council of Canada (L.Y.).

\appendix

\bibliographystyle{unsrt}
\bibliography{Q5ref}

\end{document}